# Screening-Limited Response of NanoBiosensors

*Pradeep R. Nair and Muhammad A. Alam.*

School of Electrical and Computer Engineering, Purdue University, West Lafayette, IN 47907, USA.

CORRESPONDING AUTHORS E-mail: pnair@purdue.edu, alam@purdue.edu

## ABSTRACT:

Despite tremendous potential of highly sensitive electronic detection of bio-molecules by nanoscale biosensors for genomics and proteomic applications, many aspects of experimentally observed sensor response (S) are unexplained within consistent theoretical frameworks of kinetic response or electrical screening. In this paper, we combine analytic solutions of Poisson-Boltzmann and reaction-diffusion equations to show that the electrostatic screening within an ionic environment limits the response of nanobiosensor such that $S(t) \sim c_1 \left( ln(\rho_0) - \frac{ln(I_0)}{2} + \frac{ln(t)}{D_F} + [pH] \right) + c_2$ where $c_i$ are geometry-dependent constants, $\rho_0$ is the concentration of target molecules, $I_0$ the salt concentration, and $D_F$ the fractal dimension of sensor surface. Our analysis provides a coherent theoretical interpretation of wide variety of puzzling experimental data that have so far defied intuitive explanation.

## I. INTRODUCTION

Nanoscale devices have recently been explored for label free, electrical detection of bio-molecules. Functionalized Silicon Nanowire (Si-NW) devices have been used to demonstrate detection of DNA [1-3] and proteins [4-5] at very low concentrations (Fig. 1a). It is generally believed that with improvement in functionalization schemes and scaling of device dimensions, multiplexed detection of bio-molecules at very low concentrations and in rapid flux can be achieved. Although tremendous improvements in sensitivity have been reported in electrical detection of bio-molecules, many aspects of experimental results are still not well explained within consistent theoretical framework.

The gap between existing theoretical understanding and the reported experiments is particularly evident from the following unexplained observations:

(i) Anomalous logarithmic dependence of sensor response on target bio-molecule concentration (Fig. 2a, [1-5]), (ii) linear dependence on pH which is crucial for protein detection (Fig. 2b, [5, 6]), (iii) non-linear dependence of sensitivity on electrolyte concentration (Fig. 2c, [7]), and (iv) anomalous time-dependence of sensor response instead of the classical Langmuir-type response (Fig. 2d, [1,2]).

In this paper we provide a comprehensive theory that interprets the four above mentioned observations quantitatively and systematically. The theory is based self-consistent solution of reaction diffusion model and Poisson-Boltzmann equation, illustrating the importance of screening-limited kinetic response of nanobiosensors. We describe the model equations in Sec. II, which are then used to study the screening-limited response of nanobiosensors in Sec. III. We conclude in Sec. IV.

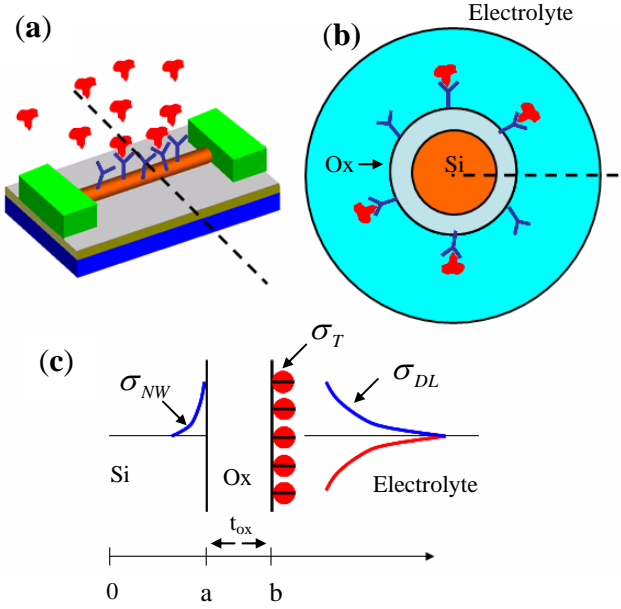

Fig. 1. Schematic of nanoscale biosensor. (a) NW surface is functionalized with receptors for target bio-molecules. (b) cross-section of the sensor along the dotten line shown in (a). (c) Charge distribution in the sensor system along the dotted line shown in (b). Device parameters are also indicated.

**A. Kinetics of Biomolecule Adsorbtion**: Diffusion-Capture model describing the kinetics of bio-molecule adsorbtion on nanonsensors (Fig. 1a) is given by

$$\frac{d\rho}{dt} = D\nabla^2\rho, \quad (1a)$$

$$\frac{dN}{dt} = k_F(N_0 - N)\rho_s - k_R N. \quad (1b)$$

Eq. (1a) represents the diffusion of target molecules to the sensor surface where $\rho$ is the concentration and $D$ is the diffusion coefficient of target bio-molecules (analyte) respectively. Eq. (1b) represents the capture of bio-molecules by the receptors functionalized on sensor surface, where $N$ is the density of conjugated receptors, $N_0$ is the density of receptors on the sensor surface, $k_F$ and $k_R$ are the capture and dissociation constants, and $\rho_s$ is the concentration of analyte particles at the sensor surface.

Based on the perturbation approach reported in Ref. [8], transport limited kinetic response of sensors (solution of Eq. (1)) is given by

$$-\frac{k_F\rho_0 + E - k_F N_{equi}}{k_F\rho_0 + k_R}\log\left(1 - \frac{N}{N_{equi}}\right) + \frac{k_F}{k_F\rho_0 + k_R}N = Et. \quad (2)$$

Here $E = N_{avo}C_D(t)/A_D$, where $N_{avo}$ is the Avogadro's constant, $C_D(t)$ the time-dependent diffusion equivalent capacitance [8], $A_D$ the area of sensor surface, and

$$N_{equi} = \frac{k_F N_0 \rho_0}{k_F\rho_0 + k_R} \quad (3a)$$

is the equilibrium concentration of conjugated molecules. It should be noted that similar solutions are also reported in [9], except that Ref. [9] considers only *steady state flux to the adsorber* while here we focus on time dynamics of molecule capture.

The two limits of Eq. (2) that will be relevant for analysis in Sec. III are: steady state (i.e., $t \to \infty$), $N \to N_{equi}$, while during the transient phase (i.e., $N \ll N_{equi}$), Eq. (2) reduces to the recently reported scaling law for nanoscale sensors [10]

$$N(t) \sim k\rho_0 t^{\left(\frac{1}{D_F}\right)}, \quad (3b)$$

where k is a geometry dependent constant, and $D_F$ is the fractal dimension of the sensor surface.

**B. Electrical Response of Bio-sensor**: The conductance modulation of a NW sensor can be derived from a simple capacitor model (see Fig. 1c). Charge conservation of the system indicates that

$$\sigma_T = -(\sigma_{DL} + \sigma_{NW}), \quad (4a)$$

where $\sigma_T$ is the charge density due to captured molecules on the sensor surface, $\sigma_{DL}$ is the net charge in the electrical double layer formed at the sensor surface (which represents the screening due to the ions in the electrolyte), and $\sigma_{NW}$ is the charge induced in the sensor. The charge density due to captured bio-molecules is given by

$$\sigma_T = \sigma_s N(t), \quad (4b)$$

where $\sigma_s$ is the charge of a bio-molecule. However, the full charge of the captured bio-molecules is not effective in modulating the conductance of sensors due to the electrostatic screening of ions present in the electrolyte. To account for screening, one must solve the non-linear Poisson-Boltzmann equation [11]

$$-\nabla^2 \Psi(r) + \kappa^2 \sinh(\beta \Psi) = q \sum_i z_i \delta(r - r_i), \quad (5)$$

where $\Psi$ is the electrostatic potential, $\beta = q/k_B T$ ($k_B$ is the Boltzmann constant, $T$ is the temperature and $q$ is the electronic charge), and $\kappa$ is Debye-Huckel parameter ($\kappa^2 = 2q^2 I_0 N_{avo} (\varepsilon_W k_B T)^{-1}$, where $I_0$ is the ion concentration in molar units and $\varepsilon_W$ is the dielectric constant of electrolyte). The *sinh* term denotes the contribution due to a 1-1 electrolyte (e.g., Na$^+$-Cl$^-$), whose ions are assumed to follow Boltzmann distribution. The right hand side denotes the fixed charge due to the bio-molecule, $z_i$ and $r_i$ denoting the partial charge and location of the atoms within the bio-molecule respectively (e.g., the phosphate ions in the backbone of a DNA strand).

The double layer charge density $\sigma_{DL}$, in terms of the potential at the sensor surface $\Psi_0$, is given by (analytical solution of Eq. (5) in cylindrical coordinates in an infinite media, without a sensor nearby [12])

$$\sigma_{DL} = -\frac{2\varepsilon_W \kappa}{\beta} \sinh(\beta \Psi_0 / 2) \left(1 + \frac{\gamma^{-2} - 1}{\cosh^2(\beta \Psi_0 / 2)}\right)^{\frac{1}{2}},$$

(6a)

where $\gamma = K_0(\kappa b)/K_1(\kappa b)$, $K$ is the modified Bessel function of second kind, and $b = a + t_{OX}$, $a$ the radius of NW and $t_{ox}$ the oxide thickness (see Fig. 1b,c). Analytic solutions of Eq. (5) are also available for other systems in literature [13, 14] and the methodology of this section can easily be extended to them as well.

For a heavily doped NW, the induced charge density is given as

$$\sigma_{NW} = -\frac{\varepsilon_{OX}}{b \log\left(1 + t_{OX}/a\right)} \Psi_0. \quad (6b)$$

Combining Eqs. (4), and (6), we have

$$\frac{\varepsilon_{OX}}{b \log\left(1 + t_{OX}/a\right)} \Psi_0 + \frac{2\varepsilon_W \kappa}{\beta} \sinh(\beta \Psi_0 / 2) \times$$

$$\left(1 + \frac{\gamma^{-2} - 1}{\cosh^2(\beta \Psi_0 / 2)}\right)^{\frac{1}{2}} = \sigma_s N(t). \quad (7)$$

The electrical response is proportional to $\sigma_{NW}$ and hence also to $\Psi_0$ (solution of Eq. (7)). Based on [15], the sensor response (relative change in conductance) is given by

$$S = \frac{\Delta G}{G_0} = \frac{2\varepsilon_{OX} \Psi_0 (N(t))}{qa^2 N_D \log\left(1 + t_{OX}/a\right)}, \quad (8)$$

where G is the conductance and $N_D$ the doping density. And Eqs. (2), (7), and (8) allow us to compute $S(t)$ self-consistently. Note that the ultimate, ideal performance limit of sensor response, as discussed in [8] is obtained if screening of target molecule approaches zero ($I_0 \to 0$ or equivalently $\kappa \to 0$), so that $S \propto \frac{(2\pi b)(\sigma_s N)}{(\pi a^2)(qN_D)}$ involves the ratio of adsorbed charge to the total doping of the NW – an intuitive result.

## II. APPLICATION OF MODEL

The self consistent screening-limited kinetic response of nanobiosensors may be obtained either by solving Eqs. (1) and (5) numerically, or by decomposing the problem into steady state and transient solution, and then solving Eqs. (2), (7), and (8) analytically. The approximate analytical solutions provide considerable insight into the problem of sensor response and are discussed next. All approximate analytical solutions

**A. Steady State Response Due to Target Molecule**: The steady state response can be obtained by replacing N(t) in the RHS of Eq. (7) with $N_{equi}$. If $k_F \rho_0 \ll 1$ (see Eq. (3a)), $N_{equi}$ can be approximated as $N_{equi} = (k_F/k_R) N_0 \rho_0$. Assuming that $\Psi_0 \gg 1/\beta$, an approximate solution of $\Psi_0$ from Eq. (7) can be obtained by neglecting the first term in Eq. (7). Using the limits $\underset{x\to\infty}{Lt}\ sinh(x) = e^x/2 = \underset{x\to\infty}{Lt}\ cosh(x)$, we find

$$\frac{2\varepsilon_W \kappa}{\beta} \sinh(\beta \Psi_0 /2) \approx \sigma_s N_{equi} = \sigma_s \frac{k_F}{k_R} N_0 \rho_0,$$

or equivalently

$$\Psi_0 = \frac{2}{\beta}\left[\ln(\rho_0) - \frac{\ln(I_0)}{2} + c_2\right], \quad (9)$$

where $c_2 = \ln\left(\frac{\sigma_s k_F N_0}{k_R}\sqrt{\frac{\beta}{q^3 \varepsilon_w N_{avo}}}\right)$. Inserting $\Psi_0$ in Eq. (8) gives the sensor response

$$S(t) = c_1\left[\ln(\rho_0) - \frac{\ln(I_0)}{2} + c_2\right], \quad (10)$$

where $c_1 = \dfrac{4\varepsilon_{ox}}{\beta q a^2 N_D \log\left(1 + t_{ox}/a\right)}$. Equation (10) shows that it is due to the inherent nonlinear screening by the electrolyte of the system, nanoscale sensors show a logarithmic dependence on the target molecule concentration $\rho_0$, and the electrolyte concentration $I_0$. *To our knowledge this is the first theoretical explanation that resolves the puzzle of log dependence of electrical response on target bio-molecule concentration [1-5] instead of the linear dependence* predicted from classical theory and observed in fluorescence measurements [9] (see the list of four puzzles in section I). Numerical solution of Eqs. (7) and (8) using $N_{equi}$ for the equilibrium concentration of captured molecules also shows the same trend (Fig. 2a), validating our analytical approach.

There are several noteworthy features in Eq. (10) and Fig. 2a: First, the slope of S(t) vs. log($\rho_0$) depends on the device parameters ($c_1$ in Eq. (10)) and any parasitic contact resistances involved. Inset of Fig. 2a (simulations) shows a comparison of the theoretical and experimental slopes of S(t) vs. log($\rho_0$). The predictions from the analytic model compares well with the experimental results, given the uncertainties in the doping density of NW devices and the effect of parasitic contact resistances (neglected in analytical model). Second, the range of analyte concentration over which the log dependence is observed can be obtained in simple terms: $N_{equi}$ becomes independent of $\rho_0$, if $k_F \rho_0 \gg k_R$. This provides an upper bound of $\rho_0$ for linear-log dependence of sensor response, i.e., $\rho_{0,max} = k_R/k_F$. Sensor response will vary logarithmically with $\rho_0$ till $\Psi_0 \gg 1/\beta$, leading to the lower bound of $\rho_0$, i.e., $\rho_{0,min} = \dfrac{2\varepsilon_W \kappa k_R e^2}{10\beta \sigma_s k_F N_0}$. It is interesting to note that while $\rho_{0,max}$ is entirely determined by the reaction coefficients, $\rho_{0,min}$ is influenced by the systems parameters as well. For typical parameters, the ratio $\rho_{0,max}/\rho_{0,min}$ predicts a range of $10^2$-$10^4$, which agrees well with the experimental data (see Fig. 2a).

**B. Steady State pH Detection**: The theoretical framework described in Sec. IIIA can be used to determine the sensor response to fluctuations in pH of the solution. The net charge density (RHS of Eq. (7)) on the NW surface can be obtained based on first order chemical kinetics of bond dissociation [16] for the particular type of surface functionalization schemes used (-OH, -NH$_2$, etc [6]). Beyond pK$_a$ ($pK_a = -\log(K_a)$, K$_a$ the dissociation constant) of the particular functionalization group, the net charge density on the sensor surface is given as $\sigma_{pH} = qN_F e^{\beta \Psi_0 + pH - pK_a}$, where N$_F$ is the density of surface functionalization groups.

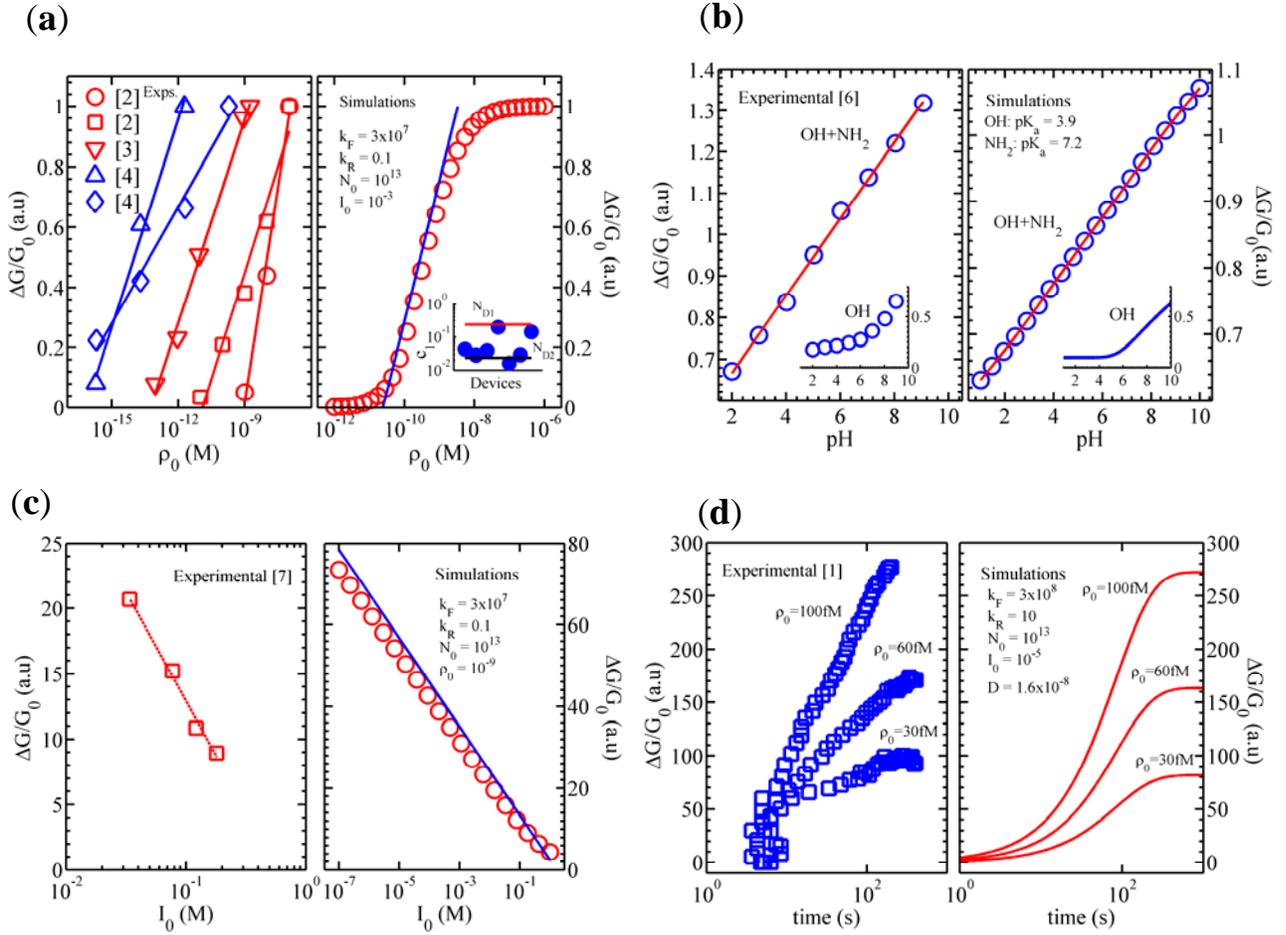

Fig. 2. Comparison between experimental (left panel) and simulated response (right panel) of nanobiosensors. (a) Variation with analyte concentration- *Logarithmic dependence on $\rho_0$*. Experimental data has been normalized. The inset in right panel shows the comparison of theoretical slope $c_1$ with experimental results ($N_{D1}=10^{19}$cm$^{-3}$, $N_{D2}=10^{20}$cm$^{-3}$. Experimental $N_D$ is of the order of $10^{19}$cm$^{-3}$). (b) Variation of sensitivity with pH (surface functionalized with amide and OH groups). - *Linear dependence on pH.* Inset shows the response with only OH groups. (c) Variation of response with ion concentration of electrolyte-*Logarithmic dependence on $I_0$*. (d) Transient response of nanobiosensors: *Logarithmic dependence on time*. The units of various simulation parameters are: $k_F$ (M$^{-1}$s$^{-1}$), $k_R$ (s$^{-1}$), $N_0$ (cm$^{-2}$), $\rho_0$ (M), $I_0$ (M), and D (cm$^{-2}$V$^{-1}$s$^{-1}$). For simulations the symbols represent numerical results while the solid lines represent analytical results.

Replacing $\sigma_s N(t)$ (RHS of Eq. (7)) by $\sigma_{pH}$ and then repeating the derivation Sec. IIIA, we find

$$S(t) = c_1 \left[ |pH - pK_a| - \frac{\ln(I_0)}{2} + c_3 \right], \quad (11)$$

where $c_3 = \ln\left( qN_F \sqrt{\dfrac{\beta}{q^3 \varepsilon_w N_{avo}}} \right)$, such that the response of sensors vary linearly with pH of solution. Numerical solution of Eqs. (7) and (8) in the presence of different surface functionalization groups (-OH and -NH$_2$), shows that linear response can be obtained over a wide range of pH (Fig. 2b, inset indicates the response in the presence of -OH groups only), in complete agreement with the reported experimental results.

*C. Steady State Response with Various Electrolyte Concentration.* Dependence of sensor response to variations in salt concentration of the electrolyte is of paramount interest in deciding the surface functionalization schemes [15]. Eq. (10) indicate that

P. R. Nair and M. A. Alam, ECE, Purdue University

for a given analyte density, $\rho_0$, the sensitivity reduces logarithmically with the ion concentration, $I_0$, also observed experimentally (see Fig. 2c). This is intuitively obvious, because screening by the ions suppresses the overall charge effective in modulating the sensor response. Numerical solution of Eqs. (7)-(8), for different $I_0$, shown in Fig. 2c, support the experimental results very well. Finally, our results indicate the critical importance of optimizing the electrolyte concentration in experiments involving biosensors, because the same target molecules may result in different sensitivity depending on the ion-concentration.

**D. Screening-Limited Transient Response of Biosensors:** The transient response can be obtained by replacing N(t) in the RHS of Eq. (7) with Eq. (3b) (i.e., when $N \ll N_{equi}$, or $T < T_{sat}$, where $T_{sat}$ is time taken for the molecule capture to reach steady state, see Eq. (2)). Following the same derivation methodology as in Sec. IIIA, we get

$$S(t) = c_1 \left[ \ln(\rho_0) + \frac{\ln(t)}{D_F} - \frac{\ln(I_0)}{2} + c_4 \right]. \quad (12)$$

Eq. (12) indicate that electrostatic screening converts the pure power-law kinetic response of biosensors (Eq. 4b) to a logarithmic dependence in time, scaled by sensor-specific fractal dimension. This is intuitively reasonable, because one can view the time-dependence of molecule arrival equivalently as a series of steady state responses of a sensor with increasing analyte concentration. Since steady-state response is logarithmic, so is the time-dependence. The predicted trends have also been observed experimentally (Fig. 2d).

## IV. DISCUSSIONS

**A. Performance Limits of Sensors:** The detection limits of nanoscale sensors in a diffusion limited regime are predicted by Eq. (12). However, screening due to the ions can significantly increase the average incubation times for achieving the same sensor response. This is evident from Eq. (12). An increase in two orders of magnitude of ion concentration (e.g., from 1mM to 0.1M), increases the average *electrical* incubation time for cylindrical NW sensors ($D_F = 1$) by one order of magnitude while for planar sensors ($D_F = $ 2) by two orders of magnitude. Hence it is important to develop functionalization schemes at low ion concentrations not only due to the electrostatic screening of ions (magnitude of sensor response) but also to reduce the time taken for obtaining a detectable signal change. Eq. (12) also implies that due to electrostatic screening, the incubation times scale exponentially with device parameters.

**B. Extraction of Reaction Coefficients:** Response of nanoscale sensors has often been used to extract the reaction coefficients $k_F$ and $k_R$ by fitting the observed experimental response with the solution of Eq. (1b) without accounting for the diffusion induced delay [2] or the nonlinear transformation between the density of captured molecules and the measured signal. Our theory, for the first time, provides a self-consistent method of characterizing the surface reaction coefficients from the electrical response of a sensor in a diffusion limited regime (using Eqs. (2), (7) - (8)).

It should be noted that the origin of logarithmic dependencies of sensor response on various parameters like $\rho_0$, $I_0$, time, and the linear dependence on pH are due to the fact that the surface potential $\Psi_0$ is such that $\Psi_0 \gg 1/\beta$. If the number or the net charge of bio-molecules captured on the sensor surface is very small, as per Eqs. (2), (7), and (8), the sensor response is expected to vary linearly with each of these parameters.

To summarize, the theory of nanoscale biosensor response, based on analytic solutions of Poisson-Boltzmann and reaction-diffusion equations, is discussed. Specifically, our model predicts that the sensor response varies (a) logarithmically with target bio-molecule concentration, (b) linearly with pH, (c) logarithmically with the electrolyte concentration, and (d) the transient response varies logarithmically with time. The theoretical predictions agree well with experimental results and have important implications for the design and optimization of nanoscale biosensors.

*Acknowledgment*: This work was supported by funds from the Network for Computational Nanotechnology (NCN) and National Institute of Health (NIH).